\documentclass[sigconf]{acmart}
\AtBeginDocument{%
  \providecommand\BibTeX{{%
    \normalfont B\kern-0.5em{\scshape i\kern-0.25em b}\kern-0.8em\TeX}}}

\setcopyright{acmcopyright}
\copyrightyear{2022}
\acmYear{2022}
\acmConference[ICSE '22]{ICSE '22: International Conference on Software Engineering}{May 21--29, 2022}{Pittsburgh, PA, USA}
\acmBooktitle{ICSE '22: International Conference on Software Engineering,
  May 21--29, 2022, Pittsburgh, PA, USA}
\usepackage{graphics}

\usepackage{caption}
\usepackage{subcaption}
\usepackage{hyperref}
\begin{document}

\title{ML-Quadrat \& DriotData: A Model-Driven Engineering Tool and a Low-Code Platform for Smart IoT Services}

\author{Armin Moin}
\email{moin@in.tum.de}
\affiliation{
  \institution{Dept. of Informatics, Technical
  Univ. of Munich (TUM)}
  \country{Germany}
}

\author{Andrei Mituca}
\email{andrei.mituca@driotdata.com}
\affiliation{
	\institution{DriotData UG \\
	Munich}
	\country{Germany}
}

\author{Moharram Challenger}
\email{moharram.challenger@uantwerpen.be}
\affiliation{
	\institution{Dept. of Computer Science, Univ. of  \\
	Antwerp \& Flanders Make}
	\country{Belgium}
}

\author{Atta Badii}
\email{atta.badii@reading.ac.uk}
\affiliation{
	\institution{Dept. of Computer Science \\
		Univ. of Reading}
	\country{United Kingdom}
}

\author{Stephan G{\"u}nnemann}
\email{guennemann@in.tum.de}
\affiliation{
	\institution{Dept. of Informatics \& Munich Data Science Institute, TUM}
	\country{Germany}
}

\renewcommand{\shortauthors}{Moin et al.}

\begin{abstract}
  In this paper, we present ML-Quadrat, an open-source research prototype that is based on the Eclipse Modeling Framework (EMF) and the state of the art in the literature of Model-Driven Software Engineering (MDSE) for smart Cyber-Physical Systems (CPS) and the Internet of Things (IoT). Its envisioned users are mostly software developers who might not have deep knowledge and skills in the heterogeneous IoT platforms and the diverse Artificial Intelligence (AI) technologies, specifically regarding Machine Learning (ML). ML-Quadrat is released under the terms of the Apache 2.0 license on Github\footnote{ \href{https://github.com/arminmoin/ML-Quadrat}{https://github.com/arminmoin/ML-Quadrat}}. Additionally, we demonstrate an early tool prototype of DriotData, a web-based Low-Code platform targeting citizen data scientists and citizen/end-user software developers. DriotData exploits and adopts ML-Quadrat in the industry by offering an extended version of it as a subscription-based service to companies, mainly Small- and Medium-Sized Enterprises (SME). The current preliminary version of DriotData has three web-based model editors: text-based, tree-/form-based and diagram-based. The latter is designed for domain experts in the problem or use case domains (namely the IoT vertical domains) who might not have knowledge and skills in the field of IT. Finally, a short video demonstrating the tools is available on YouTube: \href{https://youtu.be/VAuz25w0a5k}{https://youtu.be/VAuz25w0a5k}.
\end{abstract}





\maketitle

\section{Introduction}
In line with the Computer-Aided Design (CAD) trend, e.g., for the hardware products, the idea of Computer-Aided Software Engineering (CASE) was proposed in 1968 mostly through the \textit{Information System Design and Optimization System (ISDOS)} project \cite{Fortin1973} at the University of Michigan, USA. Later in 1971, Softlab GmbH, based in Munich, Germany brought the world first commercial Integrated Development Environment (IDE), called \textit{Maestro I} to the market. Ever since, numerous software products and services in the forms of tools, workbenches and integrated environments have aimed at increasing the productivity of software development and the quality of software systems. The examples at the present time, include, but are not limited to the Eclipse Java Development Tools (JDT), the Apache NetBeans, and the Jupyter Notebook, which is a modern web-based interactive environment that is widely used by data scientists. Given the close ties between the software systems, the Internet, which is being transformed into the Internet of Things (IoT), as well as Data Analytics and Machine Learning (DAML) today, those CASE tools that could enable both Automated Software Engineering (ASE) and Automated Machine Learning (AutoML) in an integrated manner for creating smart IoT services would be desired. 

In this paper, we present ML-Quadrat \cite{ML-Quadrat, Moin+2022-SoSyM} that is an innovative CASE tool prototype, built on top of ThingML \cite{ThingML, Harrand+2016}, an open-source project that is based on the Eclipse Modeling Framework (EMF) and the Xtext framework. Similar to ThingML, ML-Quadrat implies the Model-Driven Software Engineering (MDSE) paradigm, specifically the Domain-Specific Modeling (DSM) methodology \cite{KellyTolvanen2008} for its users. Therefore, it enables automated source code generation out of the software models for the entire software solution of smart services for Cyber-Physical Systems (CPS) and the IoT. Unlike ThingML, the MDSE practitioners using ML-Quadrat gain access to the APIs of the libraries and frameworks for ML at the modeling level. This means, using the enhanced software models in ML-Quadrat, the provided model-to-code transformations (i.e., code generators) become capable of producing not only the source code, but also the ML models for the target IoT solutions. Further, the generated source code is able to train the ML models, deploy and use them as required, and possibly re-train them by observing new data samples later. 

The contribution of this paper is twofold: (i) It demonstrates the novel, open-source, research prototype of ML-Quadrat. (ii) It demonstrates the early prototype of DriotData, a web-based Low-Code platform that adopts and exploits ML-Quadrat in the industry to enable a different user group, namely citizen data scientists and citizen/end-user software developers. These are usually subject matter experts (e.g., presales engineers) who work at companies that are not IT companies, but require IT expertise to create and offer digital services, based on the IoT and AI to their customers.

The rest of this paper is structured as follows: Section \ref{sota} reviews the state of the art briefly. Moreover, ML-Quadrat and DriotData are presented in Section \ref{ml2-dd}. Finally, Section \ref{conclusion} concludes.

\section{State of the Art}\label{sota}
In this section, we briefly review the related work in the literature and categorize them into four clusters. First, there exist various domain-specific MDSE tools serving different vertical domains, e.g., for the design, verification and implementation of embedded systems in the safety-critical applications in the automotive and \linebreak aerospace industries. Examples include, but are not limited to the \textit{MATLAB / Simulink} product of MathWorks, the products of dSpace, e.g., \textit{TargetLink}, as well as the open-source tool \textit{AutoFOCUS} \cite{Aravantinos+2015}. In the CPS/IoT domain, \textit{ThingML} \cite{ThingML, Harrand+2016} and \textit{HEADS} \cite{HEADS} (that is based on ThingML) are the state-of-the-art open-source solutions for creating the heterogeneous and distributed IoT services. What was missing in this landscape was an out-of-the-box support for AI, specifically DAML. This is provided by ML-Quadrat \cite{ML-Quadrat, Moin+2022-SoSyM, Moin+2018, Moin+2020}. All of the said tools focused on the full source code generation, not only generating the skeleton of the code.

Second, the MDSE paradigm is also applied to the Machine Learning (ML) Domain. Bishop \cite{Bishop2013} proposed this idea for the first time, and provided the open-source tool for Probabilistic Programming, called \textit{Infer.NET} \cite{InferNet}. Unlike ML-Quadrat that enhances software models to become capable of generating ML models and dealing with them, Infer.NET used a specific type of ML models, namely Probabilistic Graphical Models (PGM) to generate the source code of the entire applications out of them. However, this approach has a number of shortcomings. First, PGMs were not expressive enough to let the practitioners model the entire CPS/IoT services. Second, they only supported code generation in C\#, whereas other programming languages, such as C, Java and Python are required for CPS/IoT as well. Finally, currently, other ML model architectures rather than PGMs, e.g., deep Artificial Neural Networks (ANN) are widely used in the industry.

The third category comprises DAML workflow designers and AutoML tools/platforms. The designer tools/platforms for DAML workflows/pipelines, such as \textit{KNIME}, \textit{RapidMiner} and \textit{Tableau} also provide partial code generation capabilities. However, these are specific to the DAML practices (and in the case of Tableau, focused on business intelligence), thus it is not possible to model the entire software solution for the smart CPS/IoT services and generate the full implementations out of the models. Furthermore, certain AutoML functionalities are already provided to some extent by them. However, there also exist other platforms, such as DataRobot and Datameer that are concentrated on AutoML services. Unlike all of them, ML-Quadrat is not specific to DAML, but aims to generate the source code and the ML models for the entire solution, including the DAML part and the rest of the IoT services, in an integrated and seamless manner. Last but not least, in contrast to ML-Quadart, the above-mentioned workflow designers and AutoML platforms did not adopt the MDSE paradigm.

Finally, over the past decade, visual programming and end-user programming paradigms for creating the Web 2.0/3.0 and IoT-based services have gained interest. Initially, the web-based, visual data / service / web mash-up creation tools, such as \textit{Yahoo! Pipes} became popular. They enabled end-users to quickly combine multiple sources of data and existing web services to create new web applications/services. Today, the so-called Low-Code IoT platforms offer similar and more enhanced features for creating new IoT services. For instance, the big players in the IT sector, including Amazon, Microsoft and IBM provide such platforms as services (Platform-as-a-Service, PaaS) in addition to their Infrastructure-as-a-Service (IaaS) business models. They also support AI/ML, and offer code generation functionalities. Another example for a Low-Code IoT platform with ML support is \textit{Waylay.io}. However, none of them followed the MDSE paradigm. 

\section{The Demonstrated Tools}\label{ml2-dd}

\subsection{The Envisioned Users}\label{envisioned-users}
The envisioned user group for ML-Quadrat is software developers who might not have extensive knowledge and skills with respect to the heterogeneous IoT hardware/software platforms and the diverse AI/DAML technologies. For instance, creating a single IoT service might require familiarity with various programming languages (e.g., C, Java and Python), operating systems (e.g., Linux and ContikiOS), micro-controllers with different hardware architectures, capabilities and instruction sets, as well as communication protocols (e.g., HTTP, MQTT and CoAP). In addition, creating smart IoT services with ML capabilities requires DAML skills and familiarity with ML libraries and frameworks, such as Scikit-Learn and Keras/TensorFlow. However, using ML-Quadrat, the practitioner does not need to be familiar with the said libraries and frameworks. The model-to-code transformations that are already developed for the different platforms, libraries and protocols, will take care of the full source code generation with the respective APIs.

Moreover, the envisioned user group for DriotData is citizen data scientists and citizen/end-user software developers who are domain experts (subject-matter experts) in their problem domains (namely, their vertical IoT use case domains), but not necessarily experts in the solution domains (e.g., IT/Software Engineering (SE)/AI/ML).

\subsection{Challenges and Hypotheses}\label{challenges-hypotheses}
The main SE challenge that we expect ML-Quadrat to address concerns the heterogeneity of the IoT hardware and software platforms, the diversity of the ML libraries, frameworks, techniques, models and algorithms, as well as the specific domain knowledge that is required for efficient DAML practices, particularly for analytics modeling, i.e., building data analytics models. The latter is obviously different from the typical SE expertise of software developers. The mentioned challenges make software development for smart CPS/IoT services very costly. Therefore, the underlying hypothesis regarding ML-Quadrat is that it can help its above-mentioned target user group, i.e., software developers in developing smart IoT services in a more efficient manner, even without any deep knowledge and skills in the field of DAML. In particular, they will get the support they need to create the smart IoT services in a shorter time and with a more satisfactory experience, than would be the case as compared to the state of the art and/or manual software development.

In the case of the DriotData prototype, the expected challenges to be addressed and the resulting underlying hypothesis for the Low-Code platform are the same as above. However, since the envisioned user group is a different one, and also the model editors will have graphical views/diagrams too, separate validation studies will be required to assess the underlying hypothesis.

\subsection{Methodologies Implied for Users}\label{methodologies-implied}
The adopted programming paradigms are MDSE (particularly, the DSM methodology with full code generation \cite{KellyTolvanen2008}) and event-driven programming. The latter is a natural fit for the reactive and interactive IoT devices. Moreover, the specific software development methodology is adopted from the prior work, namely ThingML \cite{ThingML} (and HEADS \cite{HEADS} that is based on ThingML). ML-Quadrat offers a desktop modeling tool in the Eclipse IDE. The user of ML-Quadrat can use either the Xtext-based textual model editor or the EMF tree-/form-based model editor to create a valid and complete software model instance for the desired, target IoT service that needs to be generated. The textual model editor offers syntax highlighting, auto-completion, as well as a number of hints and tips at the design-time.

\begin{figure}
\centering
\includegraphics[width=0.75\linewidth]{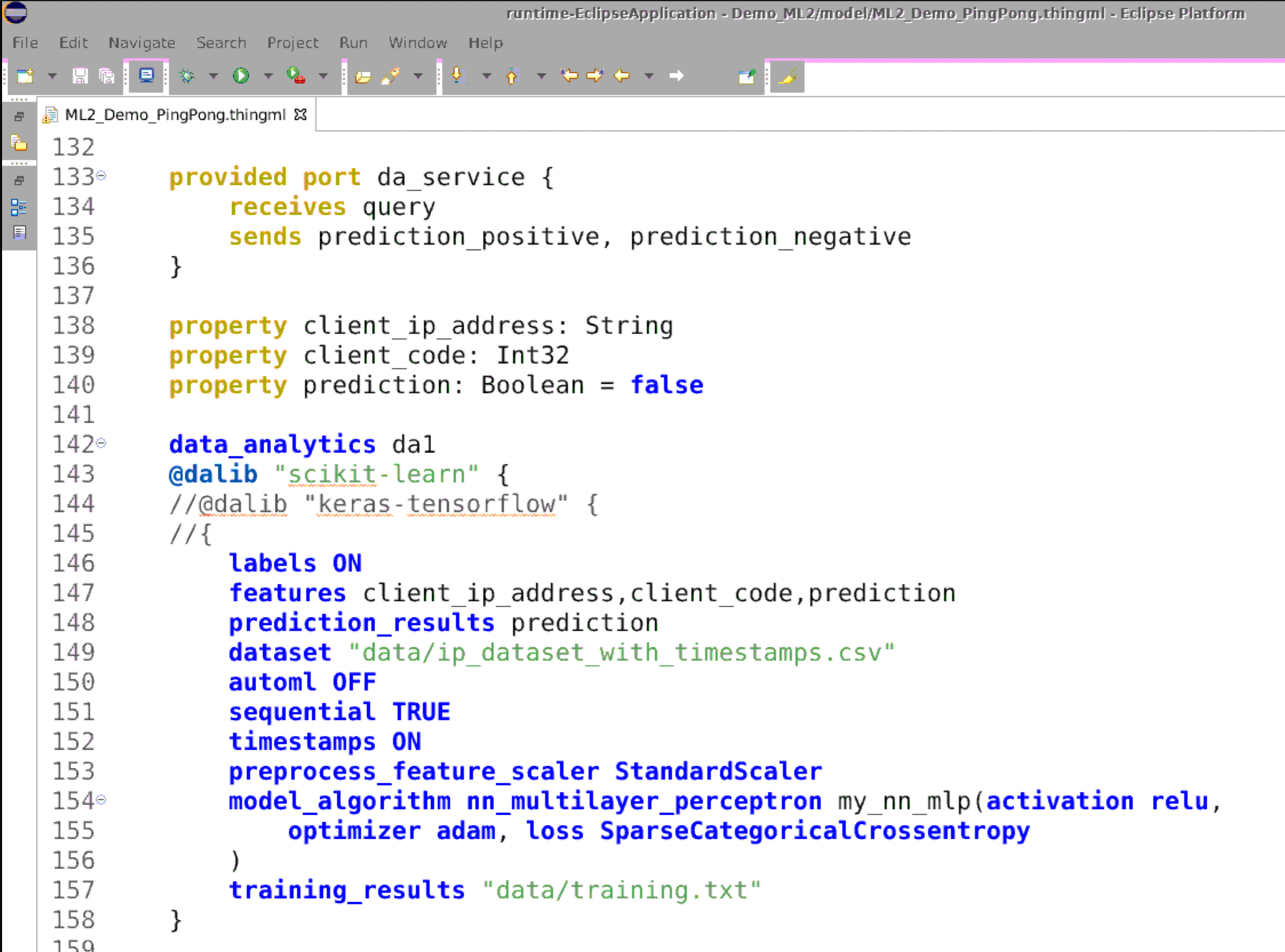}
\caption{The text-based model editor in the desktop version of ML-Quadrat that is based on the EMF and Xtext.}
\label{fig:ml2-desktop-text}
\end{figure}

\begin{figure}
\centering
\includegraphics[width=0.7\linewidth]{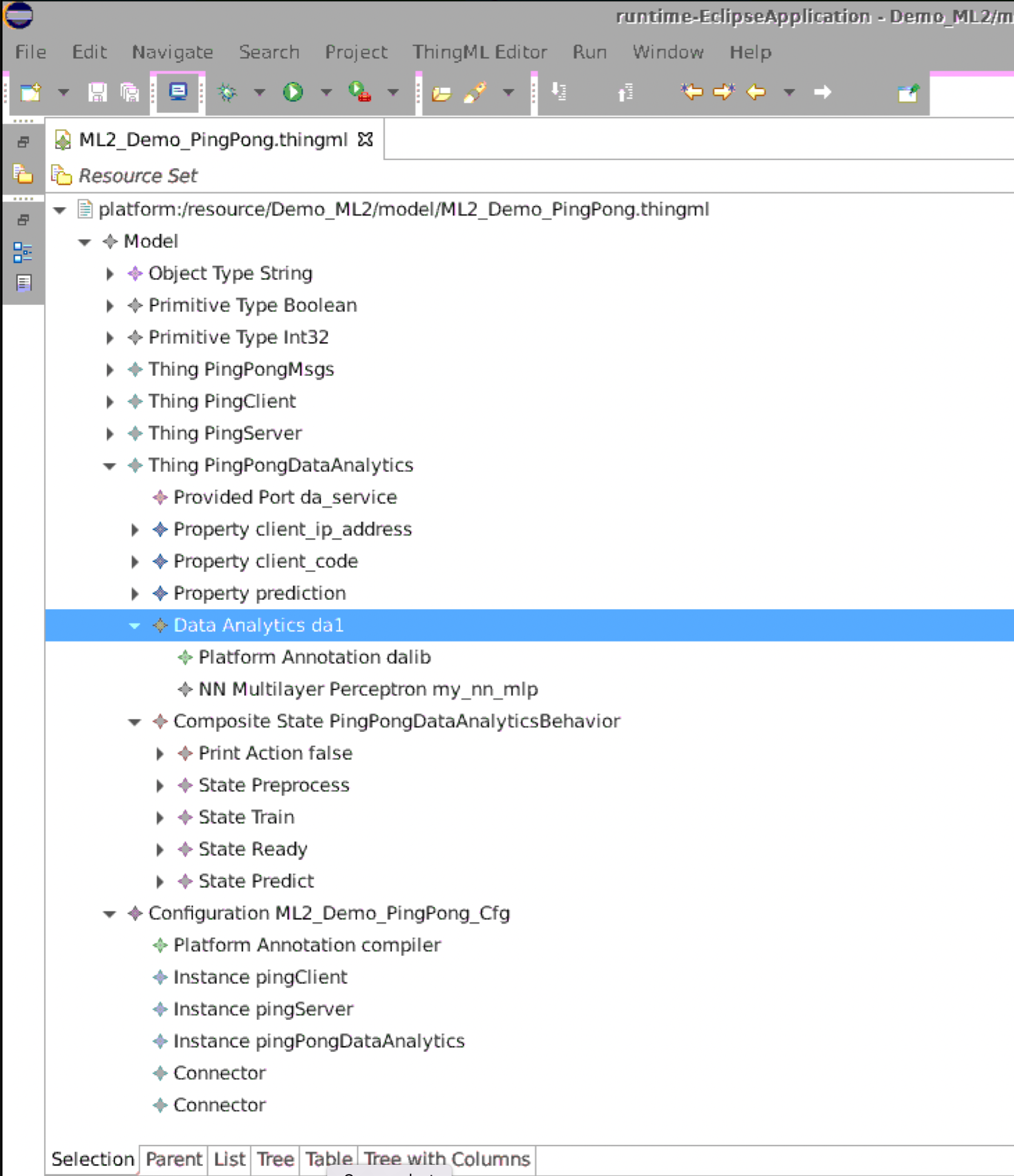}
\caption{The tree-/form-based model editor in the desktop version of ML-Quadrat that is based on the EMF.}
\label{fig:ml2-desktop-tree}
\end{figure}

\begin{figure*}
\centering
\includegraphics[width=0.65\linewidth]{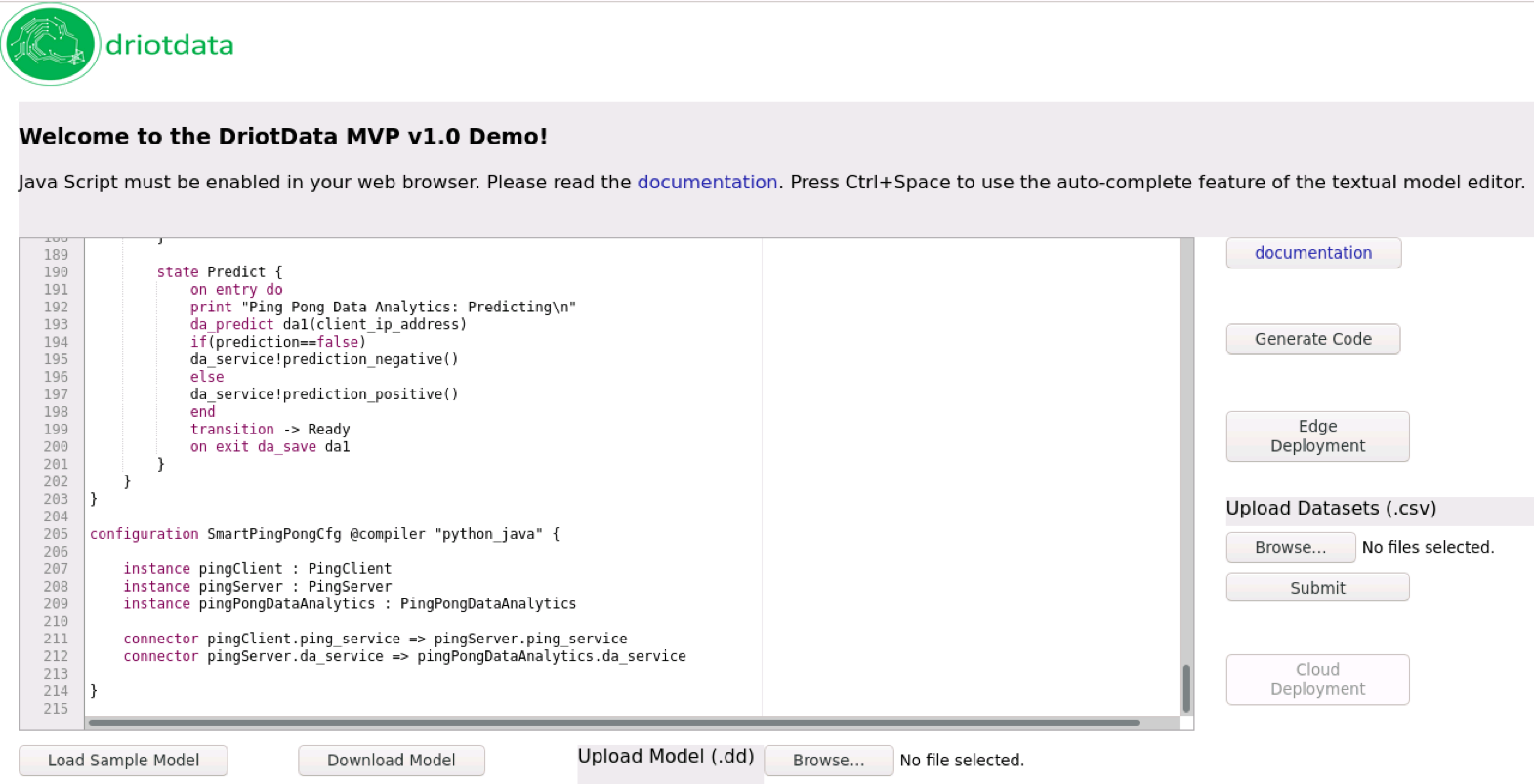}
\caption{DriotData v1.0}
\label{fig:dd1}
\end{figure*}

\begin{figure*}
\centering
\includegraphics[width=0.65\linewidth]{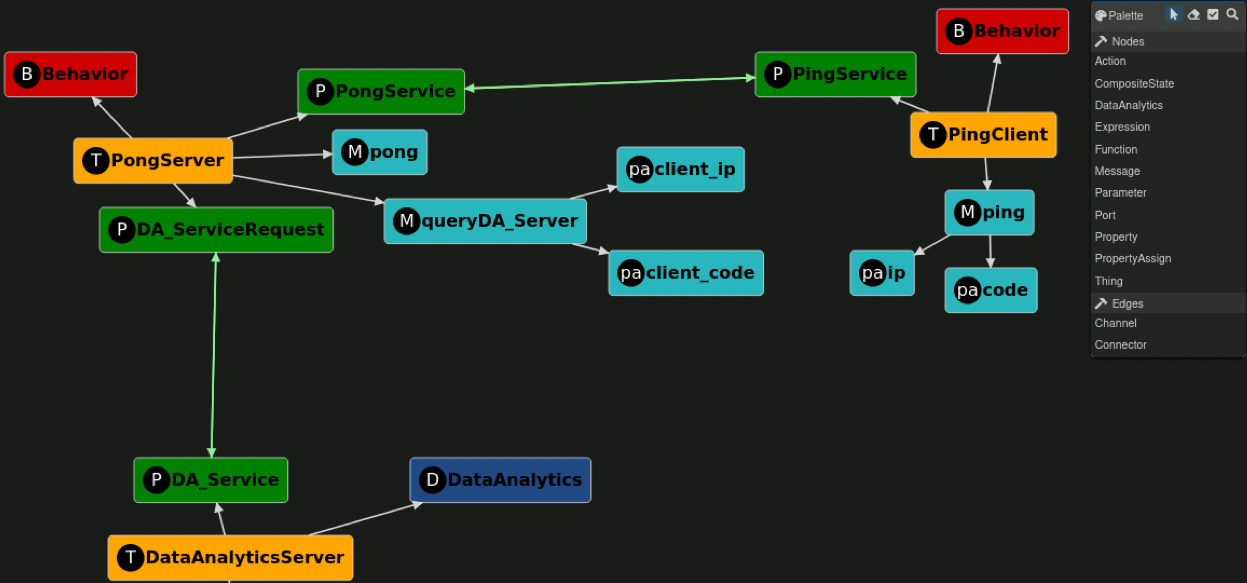}
\caption{DriotData v2.0 (under development)}
\label{fig:dd2}
\end{figure*}

\begin{figure*}
\centering
    \begin{subfigure}{.7\textwidth}
	    \centering
	    \includegraphics[width=0.85\linewidth]{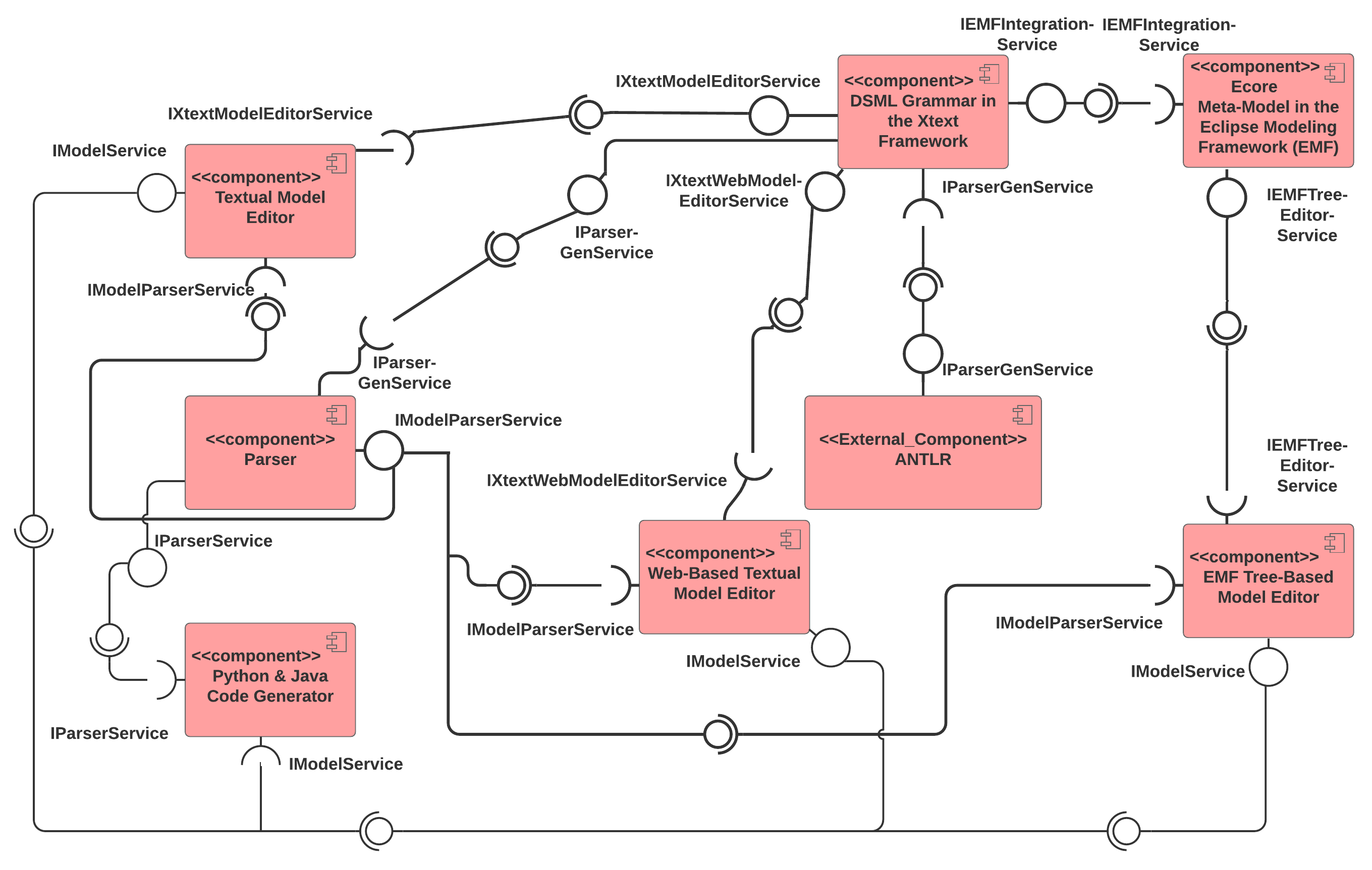}
	    \caption{The overall architecture of DriotData v1.0}
	    \label{fig:architecture}
	\end{subfigure}
    \begin{subfigure}{.2\textwidth}
        \centering
	    \includegraphics[width=0.85\linewidth]{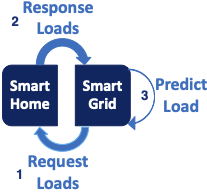}
	    \caption{A sample generated solution for the validation case study in Section \ref{conducted-study}: The smart grid must make predictions for the missing values in the received power loads of electrical appliances.}
	    \label{fig:CaseStudy}
    \end{subfigure}
\caption{}
\label{fig:architecture-casestudy}
\end{figure*}

However, the DriotData prototype comprises two versions. The first version (v1.0), which is used for the current validation study that is reported in Section \ref{conducted-study}, utilizes the Xtext web integration and the Java Servlets technology to provide the textual model editor of ML-Quadrat, as well as a number of other features through a web-based, in-browser tool. This is more convenient than the desktop version for the users since they do not need any software installation and the code generation can also be performed at the click of a button in their browser. The code will be generated on our server and they can simply download the generated code that includes the build scripts too. Additionally, the second version (v2.0) of the DriotData prototype, which is still in development, offers a diagram-based and a tree-/form-based model editor. We demonstrate ML-Quadrat and both versions of DriotData in the supplementary video\footnote{\href{https://youtu.be/VAuz25w0a5k}{https://youtu.be/VAuz25w0a5k}}. Also, Figures \ref{fig:ml2-desktop-text}, \ref{fig:ml2-desktop-tree}, \ref{fig:dd1} and \ref{fig:dd2} illustrate the text-based model editor in the desktop version of ML-Quadrat, the tree-/form-based model editor in the desktop version of ML-Quadrat, the web-based DriotData v1.0 prototype, as well as the web-based DriotData v2.0 prototype (under development), respectively. Further, the overall architecture of the DriotData prototype v1.0 is depicted using the UML Component diagram notation in Figure \ref{fig:architecture}.

\subsection{Validation Studies}\label{conducted-study}
The Technology Readiness Level (TRL)\footnote{\href{https://ec.europa.eu/research/participants/data/ref/h2020/wp/2014_2015/annexes/h2020-wp1415-annex-g-trl\_en.pdf}{https://ec.europa.eu/research/participants/data/ref/h2020/wp/2014\_2015/annexes/h2020-wp1415-annex-g-trl\_en.pdf}} of ML-Quadrat is estimated as TRL 4 since the technology has been already validated in the laboratory. The focus was on a Proof of Concept (PoC) to confirm the feasibility of the innovative idea. For the validation, we conducted an empirical user study of the DriotData v1.0 prototype, namely the web-based version of ML-Quadrat with a balanced group of four volunteer Computer Science experts\footnote{Please note that one of them, Andrei Mituca, is also a co-author of this work.}. Here is the summary of their profiles: (i) Gender: Female, Age group: 25-39, Current position: Academic, Degree: PhD, SE skill level: High, ML skill level: High, MDSE skill level: Low, IoT/CPS skill level: Low; (ii) Gender: Male, Age group: 25-39, Current position: Academic, Degree: Master's, SE skill level: Medium, ML skill level: Low, MDSE skill level: Medium, IoT/CPS skill level: Medium; (iii) Gender: Male, Age group: 25-39, Current position: Industrial, Degree: Master's, SE skill level: Medium, ML skill level: Low, MDSE skill level: Medium, IoT/CPS skill level: Medium; (iv) Gender: Male, Age group: 25-39, Current position: Industrial, Degree: PhD, SE skill level: High, ML skill level: High, MDSE skill level: Low, IoT/CPS skill level: Low. We assigned them two tasks concerning the case study that is set out in Figure \ref{fig:CaseStudy}: This comprises of a smart home and a smart grid. The smart grid periodically requests the individual electrical power consumption loads of the electric appliances in the smart home. The idea is to check if a certain “X” guaranteed-by-design holds true where “X” here happens to be “compliance with energy efficiency operation”, i.e., to verify that agreements concerning the avoidance of using particular energy-hungry appliances at peak times traded off with incentives or rewards are being complied. The smart home responds with the requested information. However, the smart grid occasionally needs to make predictions to extrapolate or impute the missing values that can occur for various reasons, such as possible network outages, or for plausibility checks of the received information. 

It transpired that the web-based tool (DriotData v1.0) can lead to the development productivity leaps of $25\%$ and $236\%$ on average, compared to the state of the art (namely ThingML \cite{ThingML}) and the pure manual development, respectively. Last but not least, the participants rated their satisfaction and overall experience. Compared to the prior work, ThingML \cite{ThingML}, the four volunteers rated their level of satisfaction with ML-Quadrat as High, High, Medium and Medium. However, compared to the pure manual development, the ratings were High, High, Medium and Low. The expert who chose Low in the latter case emphasized the simplicity of the chosen task for them in this experiment and admitted that in a more complicated IoT scenario that involved a higher degree of heterogeneity of the edge devices, they would have less chance of completion of a full system development.

However, the conducted empirical study was just a pilot, exploratory study with a limited scope, due to the resource constraints. Therefore, it cannot be relied on for any rigorous validation or quantitative analysis. In the future, an extensive empirical evaluation using a larger group, more use cases with more challenging tasks that might involve heterogeneous IoT platforms and resource-constrained IoT devices, as well as randomized controlled experiments will be necessary. In the present work, we used convenience sampling for finding the volunteers in the authors' networks. Furthermore, we plan to conduct validation studies for the DriotData v2.0 prototype with the diagram-based and the tree-/form-based model editors, in addition to the current textual model editor, in the browser. One of the main goals of the future studies will be finding the right matches between the profiles of the users of our tool and the type of model editor that they will prefer, as well as the kind of information and the way of their presentation to each user group in each of the said model editors. For instance, how much information must be shown or hidden in the diagram-based view to avoid making it cluttered and possibly confusing for its intended audience. Finally, our preliminary, exploratory interviews with the IoT domain experts who fit into the target audience of the DriotData solution hint towards the preference of many technical experts who have some programming background and/or hardware design skills to use our existing textual model editor to design their target IoT solution rather than the planned diagram-based and/or tree-/form-based model editors. This is in accordance with the prior work that emphasized the intrinsic inclination of developers towards text-based modeling with Domain-Specific Languages (DSL) rather than visual modeling using graphical model editors of the respective DSLs (e.g., see \cite{Groenniger+2007, Harrand+2016}).

\section{Conclusion}\label{conclusion}
In this paper, we demonstrated the ML-Quadrat and DriotData prototypes. The latter exploits and adopts the former in the industry. We showed that our web-based tool, DriotData v1.0 (i.e., the web-based version of ML-Quadrat), can lead to productivity leap in software development of smart IoT services. Further, our DriotData v2.0 tool, which has a graphical, diagram-based model editor too, is still under development. Finally, we plan to conduct an extensive empirical user study in the future.

%\begin{table}
%  \caption{Frequency of Special Characters}
%  \label{tab:freq}
%  \begin{tabular}{ccl}
%    \toprule
%    Non-English or Math&Frequency&Comments\\
%    \midrule
%    \O & 1 in 1,000& For Swedish names\\
%    $\pi$ & 1 in 5& Common in math\\
%    \$ & 4 in 5 & Used in business\\
%    $\Psi^2_1$ & 1 in 40,000& Unexplained usage\\
%  \bottomrule
%\end{tabular}
%\end{table}
%
%\begin{table*}
%  \caption{Some Typical Commands}
%  \label{tab:commands}
%  \begin{tabular}{ccl}
%    \toprule
%    Command &A Number & Comments\\
%    \midrule
%    \texttt{{\char'134}author} & 100& Author \\
%    \texttt{{\char'134}table}& 300 & For tables\\
%    \texttt{{\char'134}table*}& 400& For wider tables\\
%    \bottomrule
%  \end{tabular}
%\end{table*}

\section*{Acknowledgments}
	This work was partially funded by the German Federal Ministry for Education \& Research (BMBF) through the Software Campus initiative (project ML-Quadrat), as well as the German Federal Ministry for Economic Affairs and Energy (BMWi) and the European Social Funds (ESF) through the EXIST program (grant 03EGSBY811).

\bibliographystyle{ACM-Reference-Format}
\bibliography{refs}

\end{document}